\documentclass{article}
\usepackage{INTERSPEECH2018,amsmath,graphicx}
\usepackage{multirow}
\usepackage{tipa}
\begin{document}


\title{Utterance-level Intent Recognition from Keywords}


  \name{Wenda Chen$^{1,2,3}$,Jonathan Huang$^2$, Mark Hasegawa-Johnson$^1$}
	\address{$^1$Beckman Institute, UIUC, USA\\$^2$ Intel Labs, Intel Corporation\\
	$^3$Institute for Infocomm Research, A*STAR, Singapore}
	\email{wchen113@illinois.edu,jonathan.huang@intel.com, jhasegaw@illinois.edu}
\maketitle
\begin{abstract}


This paper focuses on wake on intent (WOI) techniques for platforms with limited compute and memory. Our approach of utterance-level intent classification is based on a sequence of keywords in the utterance instead of a single fixed key phrase.  The keyword sequence is transformed into four types of input features, namely acoustics, phones, word2vec and speech2vec for individual intent learning and then fused decision making. If a wake intent is detected, it will trigger the power-costly ASR afterwards. The system is trained and tested on a newly collected internal dataset in Intel called AMIE, which will be reported in this paper for the first time. It is demonstrated that our novel technique with the representation of the key-phrases successfully achieved a noise robust intent classification in different domains including in-car human-machine communications. The wake on intent system will be low-power and low-complexity, which makes it suitable for always on operations in real life hardware-based applications.

\end{abstract}
\noindent\textbf{Index Terms}: intent recognition, always-on system, low-power listening, keyword identification,wake on intent

\section{Introduction}

  Previous solutions for keyword listening and detection systems mainly focused on the single key phrase detection such as Alexa, OK Google, Hey Cortana, etc. From a user experience point of view, these systems are somewhat unnatural, as it requires the user to always say the key phrase.  Our system wakes up on intent, so that the use of key phrase can be eliminated (or reduced, as a backup in case Wake on Keyword fails).   
  
  
  For example, in the autonomous driving scenarios where the passengers are talking to the car agent for instructions, one to three sequential keywords in an utterance may be detected to determine an intent of the speaker to utilize an ASR system. In the utterances, examples of intent determination include:
  
  1. if the first keyword is `increase' and second keyword is `speed', the corresponding intent is `faster'; 
  
  2. if the first keyword is `turning' and second keyword is `way', the corresponding intent is `destination'; 
  
  3. if the first keyword is `slowing' and second keyword is `down', the corresponding intent is `slower'.
  
 In all three examples, it is inappropriate (and dangerous!) for an autonomous vehicle to change its behavior immediately in response to a few detected keywords.  Instead, it is appropriate for the vehicle to wake up its ASR, in order to generate a careful and complete transcription of the user utterance, so that an optimal response can be computed.

  
  To achieve the goal of WOI, one possibility is to run large vocabulary ASR, but the heavy processing required makes it not practical. For example, battery powered devices such as Nexus 5 smartphone only contain a 2.26 GHz quad-core CPU and 2 GB of RAM which is far from sufficient to run the always-on GPU trained CTC/DNN based acoustic models that contains millions of parameters \cite{google16}. Our system will detect keywords relevant to the intent and use enhanced features to perform the final intent classification. After the intent is detected, it will trigger the more power-costly ASR afterwards. The WOI system will be low-power and low-complexity, which makes it suitable for always on operation with suitable hardware focusing on (ultra-low) power always listening such as Gaussian Mixture Models and Neural Networks Accelerator (GNA) and Movidius Neural Compute Stick. \cite{yzxc,1777-17,intel}.

\section{Related Work}

Keyword listening, wakeup and detection research has attracted a lot of recent effort \cite{mandal2014recent}. Commercial speech recognition engines such as Siri and Alexa are examples of such systems. Typically the system focuses on modeling specific keywords or a set of keywords in lexicon and triggers the following components if detected. But when the need of utterance level intent arises, traditional methods rely on automatic speech recognition (ASR) at the full sentence level, which is both a power-hungry solution and (in a noisy environment) potentially a less accurate solution. 
Breaking the utterances into keywords and learning the rules for intent recognition hence become crucial in these situations. 

Intent recognition from spoken words is also an important research area \cite{al2010text,tur2011intent}. Traditionally, intent recognition is based on the text only, hence requires the accurate spoken information retrieval and grammatically correct sentence construction. However, in the applications of human-machine interaction, noisy keywords spoken by the user have to be captured and used for intent inference. The system has to overcome the difficulty of not having clear intent rules from text semantics, and should therefore use machine learning techniques for intent classification.

Such methodology could be applied to different application domains where an agent is waiting for instructions from the user in an always-on mode.  Running in low power hardware, the system is able to capture the core intent from the user while saving resources on battery power devices. Such system is important in AI for human-agent communication and understanding when the environment is a noisy room or inside a car. The ‘intelligent’ meetings, parties, and in-car environment that require active AI based human-computer interactions all involve the need for always-listening low-power intent recognition systems.

The paper is organized in this way: section 3 discusses the general structure of the wake on intent system; section 4 describes the AMIE dataset we used to test our technique; section 5 discusses the specific intent classification model we are proposing and evaluating in this paper; section 6 presents the experimental methods and results with discussions; section 7 concludes the paper.



\section{General System}


The full multi-stage wakeup system we used for the research is described in Figure 1. The keyword detection model, keyword listening model and classification system model are later summarized in further details. In the WOI system, an utterance with an intent label is modeled by one to three keyphrases with optional silence in between. Given the keyword recognition and segmentation tools, we will show how the keywords from either speech or text are used in intent recognition task. If a ``wake" intent is detected, the system will trigger a power-hungry and more complicated ASR engine and text model for processing.  

The full WOI system flowchart is:

1. Self loop if the keyword is un-detected;

2. Segment the keyword and Pass to input of intent classification, if keyword detected;

3. If End of utterance detected, Make the intent decision, and Send to output.

This system is the platform developed in Intel, in which our intent recognition module will be applied.

  \begin{figure}[!htb]
  \centering
  \centerline{\includegraphics[width=7cm]{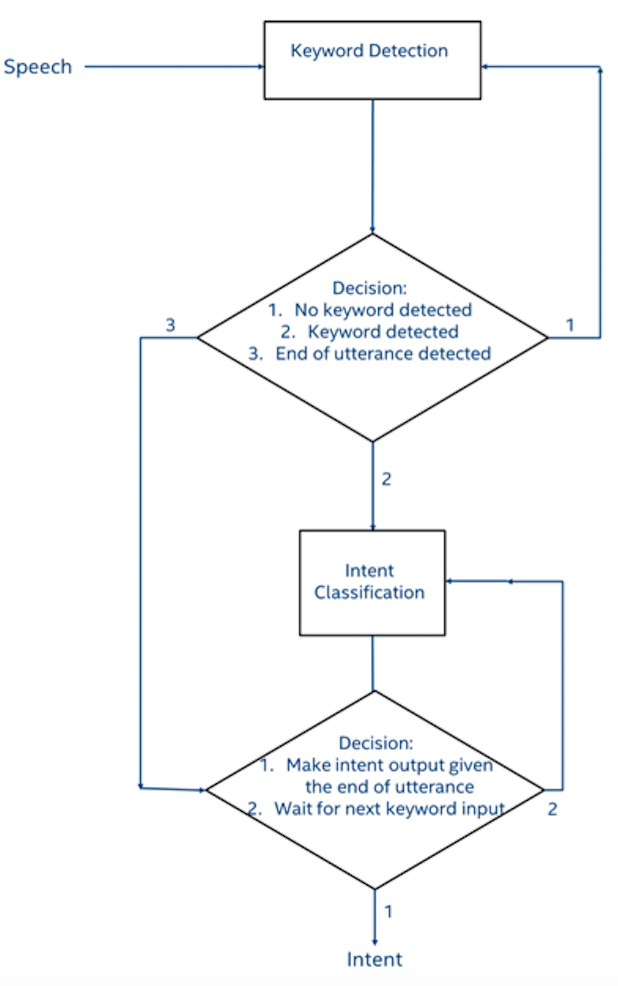}}
\caption{\textit{Full System Flowchart}}
\label{fig:prlm}
\end{figure}

 \begin{figure}[!htb]
  \centering
  \centerline{\includegraphics[width=9cm]{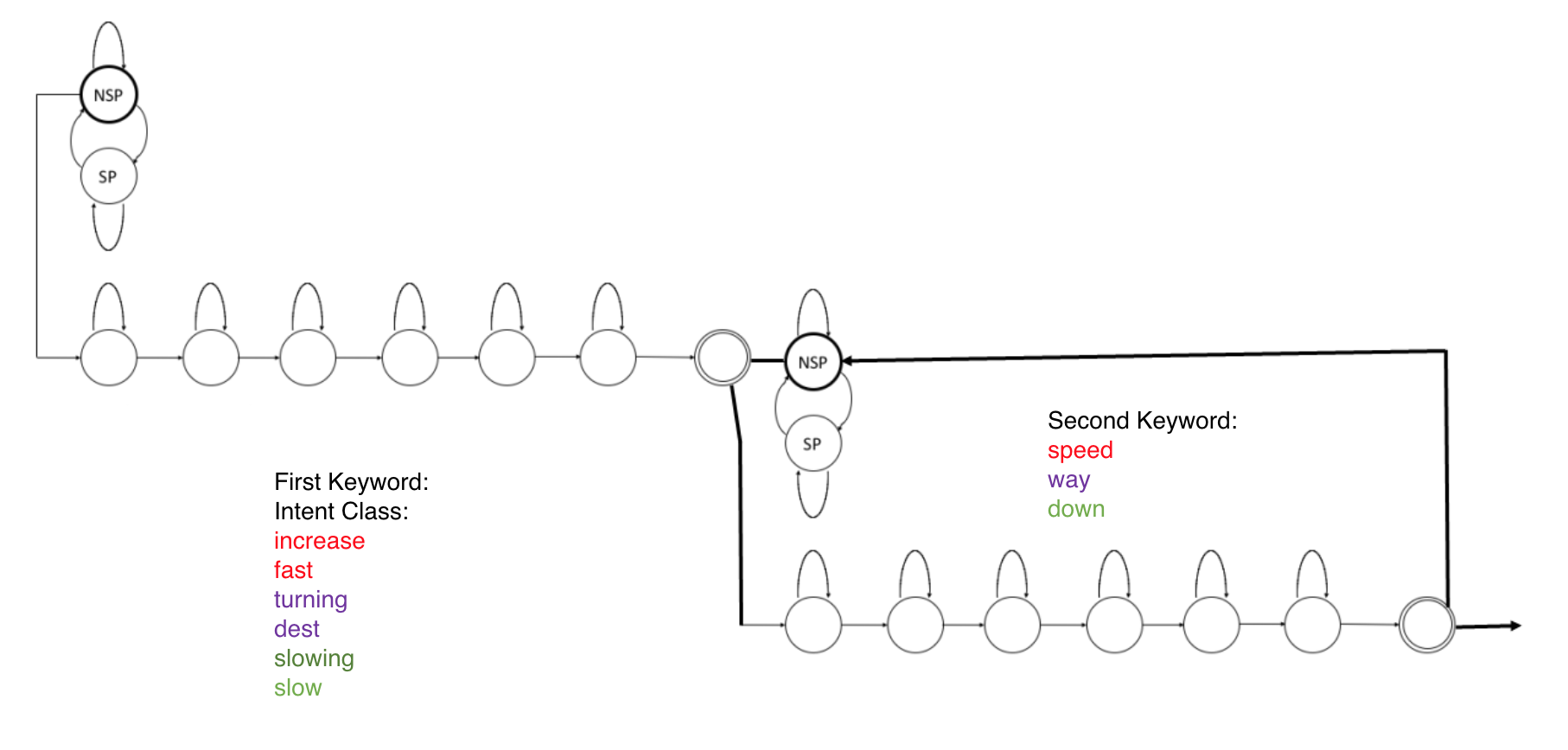}}
\caption{\textit{Keyword Listening Model}}
\label{fig:prlm}
\end{figure}

 \begin{figure}[!htb]
  \centering
  \centerline{\includegraphics[width=9cm]{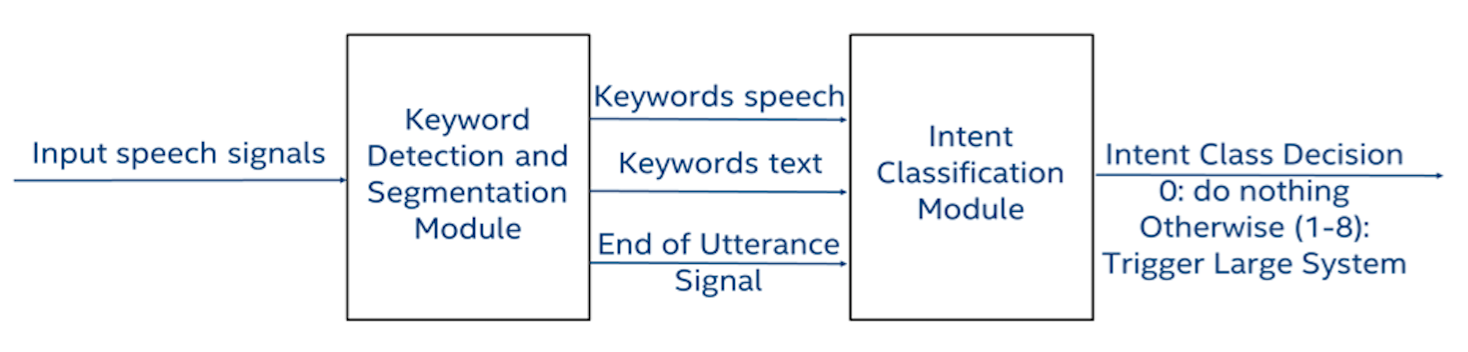}}
\caption{\textit{Classification System Model}}
\label{fig:prlm}
\end{figure}










The first stage Wakeup Word system has a rejection model representing all the silence, noises, and non-keyphrase speech. The keyphrase sequence represents subphonetic units, e.g., triphone states. For the phrase INTEL, the phoneme sequence would be IH2 N T EH1 L in ARPABET lexicon notation, the tri-phone sequence would then be
eps/IH2/N IH2/N/T N/T/EH1 T/EH1/L EH1/L/eps. In order to model the endpointing, we add a dummy state after each keyword model to represent arbitrary speech, noise, and/or silence after the keyphrase. By observing the transition from keyword model to dummy model, we can estimate precisely (with an average error of 50ms) the time-stamp when the keyphrase ends.

The keyword listening model is shown in Figure 2. The system keeps listening to the keywords and triggers the larger system once the current keyword sequence gives the utterance intent with great confidence. If one keyword is detected, and is insufficient for the detection of a ``wake" intent, and if the utterance is still not over, then the system continues waiting for the next stage (keyword) to be detected until the end of the utterance.

Classification system model is shown in Figure 3. The essential system process is: given the output of the keyword detection module and the detection of ‘end of utterance’, our follow up model for intent classification uses four features: speech2vec keyword vectors, phones, words and acoustic spectrograms. Given the keyword recognition and segmentation tools, we will show how the keywords from either speech or text are used in intent recognition task.

\section{AMIE Dataset}




The AMIE corpus, collected by Intel Labs, consists of the videos recorded for the smart cabin in-car environment where the virtual agent is talking to and receiving instructions from the passengers to take actions in the car.
This is the first time this dataset is published, while the initial research efforts during the data collection process was reported in \cite{HRI2018}. The data contains recordings of 30 subjects tasked with performing various activities inside the cabin of a real vehicle in real traffic. The activities have been designed around a game play (i.e. scavenger hunt) for one hour where real-time instructions on targets and tasks have been provided to each subject. The real-time instructions were provided by a human (called the “Game Master”) over the phone. A total of 20 rides were performed where half of the rides included a pair of subjects while the other half involved a single subject. The experience has been designed as a ``Wizard-of-Oz" simulation of an autonomous vehicle where the car is driven by a real human driver and the autonomy of the car is mimicked by a human called the “AMIE agent”. The driver is not allowed to directly communicate with the subjects while the AMIE agent interacts with the subject (i.e. the passenger) accepting commands and asking questions for disambiguation as and when needed. Some examples of the tasks performed by the subjects include eating, drinking, picking up items, entering and exiting the car, moving between locations etc.

The hardware setup used in the audio and video data collection used 5 cameras for the passengers and road, and Lapel mics for the passengers. The collected audio visual data profile is:

Video data: A total of 404 GB of data was collected over 20 sessions from 6 RGB cameras. Total video recording time is 1280 hours.

Audio data:

(i) – Lapel mic data: 135 hours of audio were captured.

(ii) – Phone conversations: Conversations between the subject and the AMIE agent provided 3 hours 45 minutes of data.

(iii) – Interviews: Each subject was interviewed before and after the ride, which yielded 9 hours of audio data.

In 20 sessions, 16 hours of dialog have been captured. These dialogues include 10590 utterances, 1466 commands/intents, 2039 responses and 744 questions. Command/intent types include “Change Destination”, “Change Route”, “Go Faster”, “Go Slower”, “Stop”, “Park”, “Pull Over”, “Drop Off”, “Open Door” as well as some outside those categories (called “Other”). Response types include “Clarify/confirm”, “Acknowledge”, “Repeat”, “Cancel” as well as some outside  those categories (called “N/A”). Question types fall into “Yes/No” and “Wh-“ categories. The manual transcribers were from AMT and GlobalME to transcribe 3347 utterances with 9 intent labels. The sample utterances with the intent labels and the total of 9 classes' labels with 3 sample keywords for each class are shown below.

Sentence 1:	you		please		go(INTENT)	to		that	destination(LOCATION)

Sentence 2:	Reroute(INTENT) 	to		three(LOCATION) 	six(LOCATION)	seven(LOCATION) 	one(LOCATION)

Class `door': open, door, get

Class `pull': get, pull, out

Class `stop': pause, do, move

Class `slow': slow, turn, slowing

Class `dest': plans, get, new

Class `park': open, door, move

Class `route': get, three, make

Class `fast': break, quickly, maximum

Class `other': trip, quickly, hold

\section{Features and Modelling}
Features to represent the keywords fall into 4 categories: MFCC sequences, phone sequence using distinctive feature vector representation for each phoneme \cite{phoible}, word level GloVe vectors sequence \cite{pennington2014glove},  and speech2vec vectors \cite{yajr}. The keyword detection system detects and segments the keywords from the utterances, then converts the speech words to the corresponding MFCC feature vector sequence, phone sequence, GloVe and speech2vec vectors. Each input feature stream is passed to a separate LSTM (Figure 4) for intent classification. The fusion of the three systems will give the final intent results. 
 \begin{figure}[!htb]
  \centering
  \centerline{\includegraphics[width=5cm]{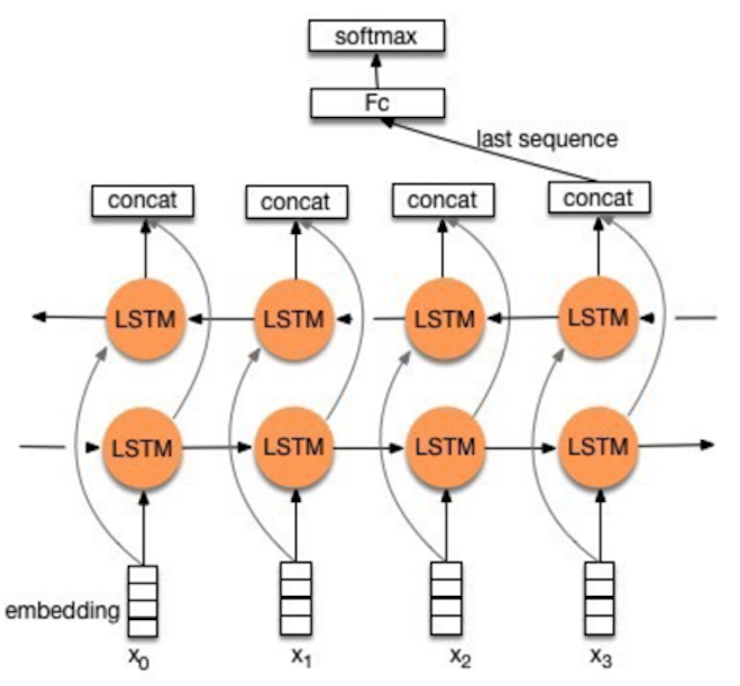}}
\caption{\textit{Individual LSTM model structure}}
\label{fig:prlm}
\end{figure}

As shown in Figure 5, system decisions are fused by a bagging mechanism, i.e., by finding the maximum output of the averaged softmax outputs over all the LSTMs.  LSTMs are trained with dropout probability 0.5; the total training time takes 2.4 hours.  The MFCC system is trained for 30 epochs, the phone-based system for 20 epochs, and the word-based systems for 15 epochs each, all with learning rate = 0.001, units per hidden layer = 20, and with 9 intent classes in the output softmax vector. The speech2vec \cite{yajr} representation takes a word-length speech segment, and converts it to an N-dimensional vector representation (we used N=100).  Training used the Librispeech corpus with a seq2seq autoencoder model \cite{panayotov2015librispeech}.










 \begin{figure}[!htb]
  \centering
  \centerline{\includegraphics[width=9cm]{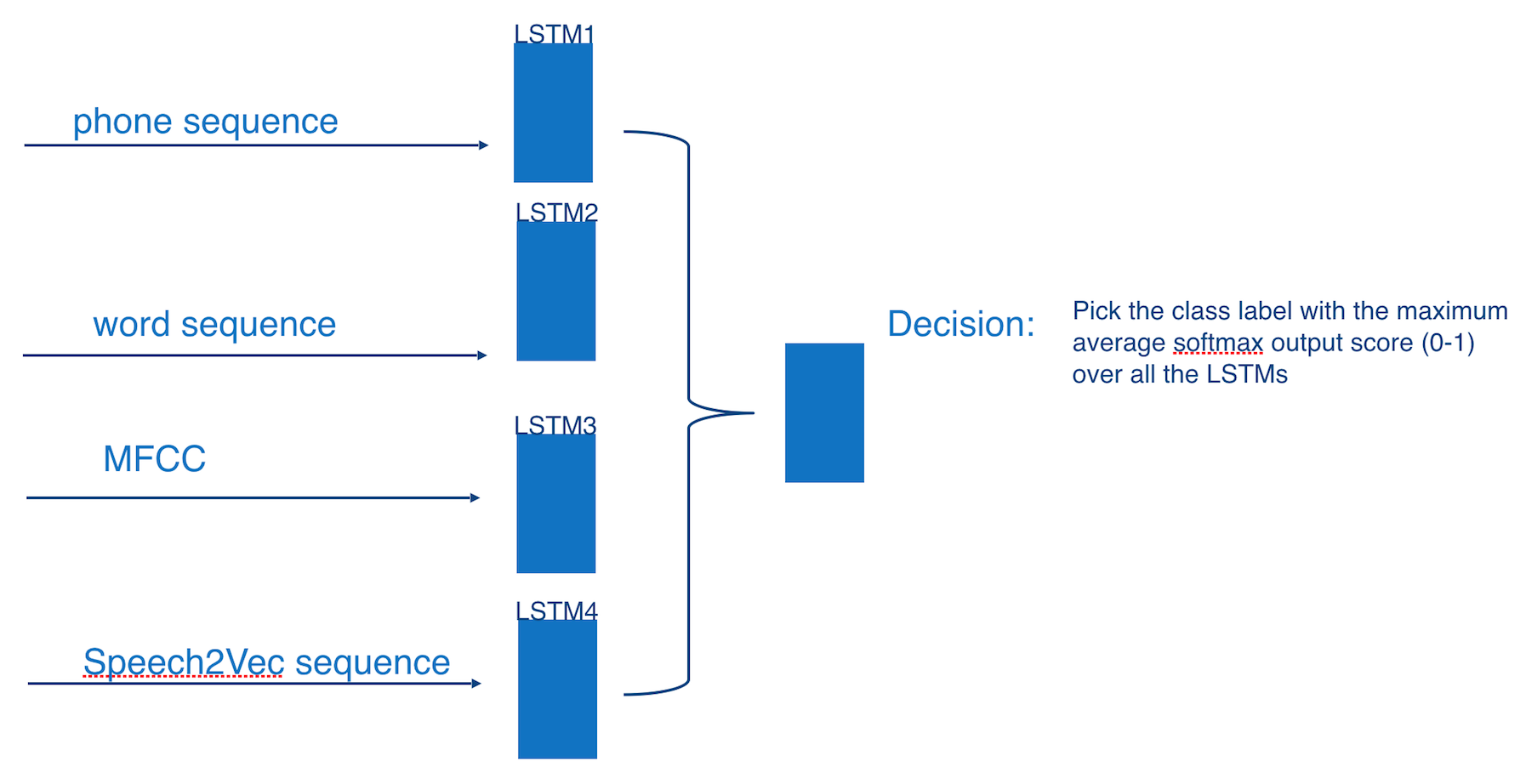}}
\caption{\textit{Intent Classification from 4 Feature Streams}}
\label{fig:prlm}
\end{figure}

\section{Experiments}
\subsection{Datasets Preparation}
We test our system on one application domain - the in-car environment (smart cabin), i.e. the AMIE corpus. 
Currently it has 1207 word-intent-labelled utterances and 380 distinct keywords-to-utterance-intent cases with 154 distinct keywords and 9 intent classes.  
We first use the Kaldi-based Gentle tool \cite{gentle17} with its own DNN acoustic model to do forced alignment for keyword segmentation and speech extraction from the utterances using both the manual transcripts and speech audios with 93\% keyword time segmentation accuracy. There are some words that were used for different intent classes but each utterance is only labeled with one intent.  

In addition to AMIE corpus, the online available keywords dataset such as Google Commands were used as well. Text-to-speech tools were also used to generate the instances of some keywords for training. 
For the speech data, the keyword dataset has also been expanded from the AMIE domain to other domains. DSTC 5 is the text corpus on intent/dialog state tracking in conversations \cite{dstc5}. The example dialog states are recommendation, explanation, acknowledgement, etc. The utterances and corresponding state labels are selected and used to expand our current corpus. 
Utterance-level intent recognition in the dialogue context is hence transferred to the new domain from the smart cabin (AMIE domain) and the features include both speech and text. 
\subsection{Methods and Results}
In Table 1, all the methods are trained and tested on the AMIE corpus with 5 fold cross validation process. The first column shows the five systems for comparison: the first row is the baseline using a commercial ASR systems, and the last row is the final fusion system. The other three rows show the systems with different input features. The second column shows the utterance level intent classification results and the third column shows the word level intent classification results.  The keyword sequence based utterance level intent classification accuracy is better than the individual word level accuracy. The fusion of the LSTM systems further improves the intent detection performance with the bagging decision mechanism as compared with the individual feature input systems.

\begin{table}[!htb]
\centering
\label{my-label}
\begin{tabular}{|c|c|c|}
\cline{1-3}
\textbf{\begin{tabular}[c]{@{}c@{}}Classification Accuracy \\for different input features\end{tabular}}  &
\textbf{\begin{tabular}[c]{@{}c@{}}Utterance level\end{tabular}}  & 
\textbf{\begin{tabular}[c]{@{}c@{}}Word level\end{tabular}}   \\ \cline{1-3}
\textbf{\begin{tabular}[c]{@{}c@{}}0. Commercial ASR \\based GloVe LSTM\end{tabular}}     &  	15\% 	&13\% \\ \cline{1-3}
\textbf{\begin{tabular}[c]{@{}c@{}}1. Forced Aligned \\MFCC LSTM\end{tabular}} 	& 80\%	&61\%\\ \cline{1-3}
\textbf{\begin{tabular}[c]{@{}c@{}}2. Forced aligned \\Phone sequence LSTM\end{tabular}}     &  	81\%&	63\%\\ \cline{1-3}
\textbf{\begin{tabular}[c]{@{}c@{}}3. Manual keyword transcript\\ based GloVe LSTM\end{tabular}} 	& 	89\%	&74\%\\ \cline{1-3}
\textbf{\begin{tabular}[c]{@{}c@{}}1+2+3: Phone+mfcc+manual\\ GloVe LSTM\end{tabular}}     &  	91\%	&75\%\\ \cline{1-3}
\end{tabular}
\caption{\textit{Intent Classification from Spoken Keywords in different features}}
\end{table}

Now we expand the domain dataset to 5 times of the size to include the added false alarm and false rejection noisy words data and the collected conversational text/speech data in the smart cabin and DSTC domains. Even with the pure text based conversations and dialogs, keyword model transfer mechanism can use acoustics from TTS signals, speech2vec, word2vec and phone sequences as the input features for the dialog state classification. Here the speech2vec was trained and obtained using Librispeech data for the text. For the case of AMIE data, speech2vec is used as an additional word level feature that assumes that the words are pronounced and recorded in standard clean English way.
\begin{table}[!htb]
\centering
\label{my-label}
\begin{tabular}{|c|c|c|c|}
\cline{1-4}
\textbf{\begin{tabular}[c]{@{}c@{}}F1 scores\end{tabular}}  &
\textbf{\begin{tabular}[c]{@{}c@{}}One Domain\end{tabular}}  & 
\textbf{\begin{tabular}[c]{@{}c@{}}Expanded\end{tabular}} & 
\textbf{\begin{tabular}[c]{@{}c@{}}Two domains\end{tabular}}  \\ \cline{1-4}
\textbf{Utterance level}     &  0.92	&0.89&	0.85\\ \cline{1-4}
\textbf{Word level} 	& 0.81&	0.76	&0.73\\ \cline{1-4}
\end{tabular}
\caption{\textit{Intent Classification from Spoken Keywords in Different Domains}}
\end{table}

The results for both individual keyword detection and combined multiple keyword detection and listening performance are compared in Table 2. 
The manual labeling of the intent classes were again obtained from AMT and GlobalME. One domain is mainly in smart cabin, expanded domain includes the added false alarm and false rejection noisy words data, and two domains are the smart cabin and DSTC. It shows that the expaned domain of data reduces the performance of the intent recognition. Table 3 shows that adding speech2vec feature will further improve intent recognition process on the two domains case.

\begin{table}[!htb]
\centering
\label{my-label}
\begin{tabular}{|c|c|c|}
\cline{1-3}
\textbf{\begin{tabular}[c]{@{}c@{}}F1 scores\end{tabular}}  &
\textbf{\begin{tabular}[c]{@{}c@{}}3 features \\(word2vec, phone, acoustics)\end{tabular}}  & 
\textbf{\begin{tabular}[c]{@{}c@{}}4 features \\(+Speech2vec)\end{tabular}}   \\ \cline{1-3}
\textbf{\begin{tabular}[c]{@{}c@{}}Utterance level\end{tabular}}     &  0.85&  0.87\\ \cline{1-3}
\textbf{\begin{tabular}[c]{@{}c@{}}Word level\end{tabular}} 	& 0.73 & 0.74\\ \cline{1-3}
\end{tabular}
\caption{\textit{Intent Classification from Spoken Keywords for Additional Features}}
\end{table}

The following experiments in Table 4 evaluate the effectiveness of all four features, namely, 1: acoustics, 2: phones, 3: word2vec, 4: speech2vec. Here each feature is passed to an LSTM based intent classification system and the final decision is made by bagging mechanism of the decision fusion from the individual systems. The testing dataset is a combination of the keywords-based utterances from the three different domains.
F1 scores show that the most effective features are word2vec and acoustics, while the relatively most ineffective feature is phones. 
In the last row, the new updated weights for each feature is determined by the relative effect of the individual features as shown the in the upper rows. Adjusting weights in this way further improves the performance of the 4 feature system.

\begin{table}[!htb]
\centering
\label{my-label}
\begin{tabular}{|c|c|c|}
\cline{1-3}
\textbf{\begin{tabular}[c]{@{}c@{}}Features/F1 scores\end{tabular}}  &
\textbf{\begin{tabular}[c]{@{}c@{}}Utterance level\end{tabular}}  & 
\textbf{\begin{tabular}[c]{@{}c@{}}Word level\end{tabular}}   \\ \cline{1-3}
\textbf{1,2,3}     &  0.85&  0.73\\ \cline{1-3}
\textbf{2,3,4} 	& 0.82 & 0.70\\ \cline{1-3}
\textbf{1,2,4}     &  0.80&  0.69\\ \cline{1-3}
\textbf{1,3,4} 	& 0.86 & 0.72\\ \cline{1-3}
\textbf{1,2,3,4}     &  0.87&  0.74\\ \cline{1-3}
\textbf{new weights}     &  0.89&  0.77\\ \cline{1-3}
\end{tabular}
\caption{\textit{Intent Classification from Spoken Keywords for Feature Combination and System Improvement}}
\end{table}



\section{Conclusion}

In this paper, we developed the intent classification model from recognized keywords. The conclusions are: 1. in a low power setting, using keyword sequences in an utterance is sufficient and necessary for utterance level intent recognition with 0.89 F1 score; 2. using the fusion of speech acoustics based features together with text input on words and phones further improves the intent detection accuracy from text only. The proposed intent recognition framework from detected keywords sets can be extended in different domains, since low power always on systems are needed for various project applications such as AMIE.

\section{Acknoweldgement}
This work was done when the first author was an intern in Intel Labs. Thanks to Cagri Tanriover in Intel for supporting and describing the AMIE dataset.
\bibliographystyle{IEEEbib}
\bibliography{refs}

\end{document}